\documentclass[pra,twocolumn,showpacs,floatfix]{revtex4}
\usepackage{graphicx,amsmath,amsfonts,amssymb,mathrsfs}
\usepackage[usenames]{color}

\begin{document}

\title{Localization of a Bose-Fermi mixture in a bichromatic  optical lattice}

\author{Yongshan Cheng$^{1,2}$\footnote{yong\_shan@163.com}
and
   S. K. Adhikari$^1$\footnote{adhikari@ift.unesp.br;
URL: www.ift.unesp.br/users/adhikari}}
\affiliation{
$^1$Instituto
de F\'{\i}sica Te\'orica, UNESP - Universidade Estadual Paulista,
01.140-070 S\~ao Paulo, S\~ao Paulo, Brazil\\
$^2$Department of Physics, Hubei Normal University, Huangshi 435002,
People's Republic of  China
 }

\date{\today}

\begin{abstract}
We
study the localization of a cigar-shaped super-fluid Bose-Fermi
mixture in a quasi-periodic bichromatic optical lattice (OL) for
inter-species attraction and intra-species
repulsion. The mixture is described by the
Gross-Pitaevskii equation for the bosons, coupled to a hydrodynamic
mean-field equation for fermions at unitarity.  We confirm
the existence of the symbiotic localized states in the Bose-Fermi
mixture and
Anderson localization of the Bose component
in the interacting Bose-Fermi mixture
on a bichromatic OL. The phase diagram in
boson and fermion numbers showing the regions of the symbiotic and
Anderson localization of the Bose component is presented. Finally,
the stability of  symbiotic and Anderson localized states is
established under small perturbations.

\end{abstract}


\maketitle

\section{Introduction}
\label{I}

Anderson localization of super-fluid
atomic gases
in weak disordered potentials with a large exponential tail
is  currently
attracting a lot of interest in both  experimental and
theoretical studies. In experimental
studies,  disorder laser
speckle \cite{billy} and quasi-periodic optical lattice (OL) \cite{NAT-453-895}
have been used to localize Bose-Einstein condensates (BEC).
The original description of Anderson localization
was based on the localization of
non-interacting quantum waves in disorder potentials due to a
cancellation of wave fronts coming from different locations of the disorder
potential. However,
super-fluid atomic gases usually are  interacting and the
study of Anderson localization has also been extended to the
case of localization of a single-component \cite{andlocus1c,inter} or
binary \cite{andlocusb} BEC under repulsive effective interactions.
Both
 quasi-periodic OL in one \cite{andlocus1c,adhisala} and three dimensions \cite{andlocus1c3d}
and random potential \cite{andlocusran} were used in these studies.
The effect of temperature on Anderson localization has also been
investigated \cite{temp}.

Another topic of current interest is the
problem of trapped binary super-fluids,
where the two components
could be two different hyperfine states of the same atom ($^{87}$Rb)
\cite{PRL-78-586} or two different atoms. In the latter case,
 degenerate
Bose-Fermi      $^{87}$Rb-$^{40}$K mixture
\cite{SCI-294-1320},  degenerate Fermi-Fermi
mixture of $^{40}$K \cite{SCI-285-1703}
and of $^6$Li
\cite{PRL-91-080406} were considered  among others.
In particular, the
Feshbach resonance technique driven by a magnetic \cite{NAT-392-151} or
optical \cite{PRL-93-123001} field allows one to vary the atomic
interaction
that opens the possibility for the  study of
localization of binary condensates with controllable interactions. A Bose-Fermi
mixture can exhibit quite distinct features as the number of atoms and
the  inter-species and intra-species interactions are varied. For
example, the intra-species repulsive interaction combined with the
inter-species attraction may give rise to the symbiotic soliton
\cite{PRA-72-033620,PLA-346-179}. In reference \cite{JPB-42-205005}, we
demonstrated that the inter-species attraction  contributes to
an attractive point-like effective potential which affects the atom
density distribution and the stability properties of the condensate
\cite{PRA-71-033609}.

Here we investigate the symbiotic localization of a cigar-shaped
Bose-Fermi superfluid mixture without a trap. The Fermi superfluid
is taken to be a mixture of equal number of spin up and down
components in a paired state. We also consider Anderson localization
of the Bose component in the Bose-Fermi mixture in a quasi-periodic
bichromatic OL when the Bose component is not localized in the
absence of the OL. In both cases the intra-species Bose interaction
is taken to be repulsive and the Fermi component is considered to be
at unitarity  \cite{jltp}. The unitarity limit is the limit of
strong attraction between spin-up and -down fermions with the
Fermi-Fermi scattering length approaching infinity. Consequently,
the properties of the Fermi superfluid in this limit become {\it
universal} and independent of the  Fermi-Fermi scattering length and
are solely determined by the Fermi energy and Fermi momentum. The
unitary limit of Fermi superfluid can now be routinely achived and
studied in laboratory \cite{jltp}. For a theoretical description of
the interacting Bose-Fermi mixture we consider the mean-field
Gross-Pitaevskii (GP) equation of bosons coupled to a mean-field
hydrodynamic density-functional equation \cite{df} for fermions at
unitarity. The present investigation involves numerical simulation
as well as analytical study based on a variational approach. The
formation of symbiotic and Anderson localizations as a function of
Bose and Fermi number  is illustrated in a phase diagram. By
numerical simulation of the coupled mean-field  model using the
split-step Fourier spectral method, we confirm the existence of the
symbiotic localized states in the Bose-Fermi mixture without
external trapping potential. We also identify Anderson localization
of the Bose component with an exponential tail assisted by the Fermi
component in the presence of a quasi-periodic bichromatic OL. We
find that the Anderson localization of the Bose component is
possible in the mixture for an inter-species attraction which
neutralizes mostly the bosonic repulsion.  Finally, we establish the
stability of both symbiotic and Anderson localizations under small
perturbations.

In Sec. \ref{II} we present a brief account of the coupled mean-field
model
and the bichromatic OL potential used in the study. The analytical
expressions for the width of the Bose and Fermi localized
states obtained by the variational analysis of the mean-field
model are in reasonable agreement of the numerical solution.
The numerical results for the static and stability
properties of the symbiotic and Anderson states
are presented  in
Sec. \ref{III}. In Sec. \ref{IIII} we present a brief summary.

\section{Analytical consideration of localization}
\label{II}

We consider a binary super-fluid Bose-Fermi mixture of
 $N_{\rm B}$ bosons of mass $m_{\rm B}$ and
$N_{\rm F}$ fermions of mass $m_{\rm F}$ at zero temperature. The
spin-half fermions are taken to be at unitarity and populated
equally in spin-up and down states. The intra-species Bose
interaction is taken to be repulsive and the inter-species
Bose-Fermi interaction is taken to be attractive. Experimentally,
this situation is, for example, accessible in a Bose-Fermi mixture \cite{SCI-285-1703, SCI-297-2240}
of bosonic $^{87}$Rb atoms in the hyperfine state $|F=2, m_{\rm
F}=2\rangle$ and fermionic $^{40}$K atoms in the two equally
populated hyperfine states $|F=9/2, m_{\rm F}=- 9/2\rangle$ and
$|F=9/2, m_{\rm F}=-7/2\rangle$. {   Theoretically,   the bosons are
treated by the usual mean-field GP equation with contact
interaction, which is equivalent to the standard hydrodynamical
equations. The fermions are treated by a mean-field
density-functional equation, which is equivalent to the standard
hydrodynamical equations for the Fermi superfluid. The bosons and
fermions are assumed to interact by a zero-range potential. Within
the framework of the density-functional theory, a coupled GP-type
equation for the mixture can be obtained \cite{df, PRA-81-053630}.
}

The system is made effectively one-dimensional (1D), assuming that
the mixture is confined in transverse directions by a tight
axisymmetric harmonic potential of frequencies $\omega_{\perp B}$
and $\omega_{\perp F}$ for bosons and fermions, respectively. The
three-dimensional equations of Ref. \cite{df} for a cigar-shaped
superfluid Bose-Fermi mixture can then be reduced to an effective 1D
form by integrating out the dependence on the radial spatial
variables \cite{PRA-81-053630}. The dynamics of the 1D mixture is
described by the dimensionless coupled time-dependent nonlinear
equations \cite{PRA-81-053630}
\begin{eqnarray}
i\frac{\partial u_{\rm B}}{\partial
t}&=&-\frac{1}{2}\frac{\partial^2 u_{\rm B}}{\partial x^2} +g_{\rm
B}|u_{\rm B}|^2u_{\rm B}\nonumber\\
&&+g_{\rm BF}N_{\rm F}|u_{\rm F}|^2u_{\rm B}+V(x)u_{\rm B},\label{model1}\\
\frac{i}{2}\frac{\partial u_{\rm F}}{\partial
t}&=&-\frac{1}{8}\frac{\partial^2 u_{\rm F}}{\partial x^2} +g_{\rm
F}|u_{\rm F}|^{4/3}u_{\rm F}\nonumber\\
&&+g_{\rm BF}N_{\rm B}|u_{\rm B}|^2u_{\rm F}+V(x)u_{\rm F},
\label{model2}
\end{eqnarray}
where $V(x)$ is the  trapping potential
acting on both the Bose and Fermi components, $u_{\rm B}\equiv
u_{\rm B}(x, t)$ and $u_{\rm F}\equiv u_{\rm F}(x, t)$ are the 1D
wave functions with normalization $\int_{-\infty}^{\infty}|u_{\rm
B}|^2 dx=\int_{-\infty}^{\infty}|u_{\rm F}|^2 dx=1$.
In Eqs. (\ref{model1}) and (\ref{model2}), we set the transverse
oscillator lengths
$a_{\perp B} = a_{\perp F}  \equiv a_\perp$
and $\omega_{\perp B} = \omega_{\perp F}  \equiv \omega_{\perp}$,
measuring length, energy  and time in
units of $a_{\perp}$, $\hbar\omega_\perp$  and $\omega_{\perp}^{-1}$, respectively
\cite{PRA-81-053630}. This implies that
$2m_F = m_B$, a condition which
is roughly satisfied by the $^{87}$Rb-$^{40}$K mixture.
  The dimensionless  interactions are
 \cite{PRA-81-053630}
$g_{\rm B}=2(a_{\rm B}/a_\perp)N_{\rm B},
g_{\rm BF}=6(a_{\rm BF}/a_\perp),
g_{\rm F}=(3\pi^2)^{2/3}(3\xi/5)N_{\rm F}^{2/3},$
where $a_{\rm B}$ is the Bose scattering length, $a_{\rm BF}$ is the
Bose-Fermi scattering length, the universal Bertsch constant $\xi$
has the value $\xi=0.4$ at unitarity \cite{RMP-80-1215,jltp}. We
consider the scattering length for collision between $^{87}$Rb atoms
to be positive, $a_{\rm B}\approx108a_0$ with $a_0$ the Bohr radius
\cite{RMP-71-463}, whereas the inter-species scattering length
between bosons ($^{87}$Rb) and fermions ($^{40}$K) is negative
$a_{\rm BF}\approx-284a_0$ \cite{SCI-297-2240, PRL-89-150403}. If
the Bosons are decoupled from the Fermions (by considering $a_{\rm
BF}=0$), they satisfy the GP equation and  it was found in Ref.
\cite{inter} that Anderson localization is destroyed due to
Bose-Bose repulsion while the stationary   states are exponentially
localized \cite{spenser}. Once a Bose-Fermi attraction above a critical value   
is introduced via a nonzero negative $a_{\rm BF}$, so
that the Bose-Bose repulsion in the GP equation is nearly
compensated for by interatomic attraction, localization is restored.

The coupled equations   (\ref{model1}) and
(\ref{model2}) will be  used  to study
 a cigar-shaped, localized Bose-Fermi system
\cite{PRL-75-854,PRA-73-053608,PRA-76-043626}. We
study  the symbiotic
localized states of this system
and Anderson localization of bosons in the
Bose-Fermi mixture. To study the symbiotic localized states
due to the strong inter-species attraction  we take $V(x)=0$.
To study the Anderson
localization, following the experiment of Roati {\it et al.}
\cite{NAT-453-895}, the potential $V(x)$ is taken to be a
quasi-periodic bichromatic OL of incommensurate wave lengths:
\begin{eqnarray}\label{pot}
V(x)=\sum_{l=1}^2 A_l \sin^2(k_lx),
\end{eqnarray}
with $A_l=2\pi^2 s_l/\lambda_l^2, (l=1,2)$, where $\lambda_l$'s are
the wavelengths of the OL potentials, $s_l$'s are their intensities,
and $k_l=2\pi/\lambda_l$ the corresponding wave numbers. Without
losing generality, we take  $s_1=10$,
$s_2=0.3s_1$,  $\lambda_1=10$ and
$k_2/k_1=(\sqrt{5}-1)/2$ \cite{NJP-11-033023}
 which roughly represent the generic
experimental  situation \cite{NAT-453-895}.

To obtain the stationary localized states of the coupled  equations
 (\ref{model1}) and (\ref{model2}), we may set $u_{\rm
B,F}(x, t)=\phi_{\rm B,F}(x)\exp(-i\mu_{\rm B,F}t)$ with $\mu_{\rm
B,F}$ the respective chemical potentials. The real wave functions,
$\phi_{\rm B,F}(x)$, obey the stationary equation,
\begin{eqnarray}
\mu_{\rm B}\phi_{\rm B}&=&-{\phi_{\rm B}''}/{2} +g_{\rm
B}\phi_{\rm B}^3+g_{\rm BF}N_{\rm F}\phi_{\rm F}^2\phi_{\rm
B}+V\phi_{\rm B},\label{sta1}\\
{\mu_{\rm F}}\phi_{\rm F}/2&=&-{\phi_{\rm F}''}/{8}+g_{\rm
F}\phi_{\rm F}^{7/3}+g_{\rm BF}N_{\rm B}\phi_{\rm B}^2\phi_{\rm
F}+V\phi_{\rm F}, \label{sta2}
\end{eqnarray}
where the prime denotes space derivative.
The stationary localized states can be
investigated by  the Gaussian  variational approach \cite{var}.
In this approach, the Lagrangian density for Eqs. (\ref{sta1}) and
(\ref{sta2}) is
\begin{eqnarray}\label{den}
\mathscr{L}&=&N_{\rm B}[\mu_{\rm B}\phi_{\rm
B}^2-{(\phi_{\rm B}')^2}/{2}-{g_{\rm
B}}\phi_{\rm B}^4/2-V\phi_{\rm B}^2] \nonumber\\
&+&N_{\rm F}[{\mu_{\rm F}}\phi_{\rm
F}^2/2-{(\phi_{\rm F}')^2}/{8}-{3}g_{\rm
F}\phi_{\rm F}^{10/3}/5-V\phi_{\rm F}^2]\nonumber\\
&&-g_{\rm BF}N_{\rm B}N_{\rm F}\phi_{\rm B}^2\phi_{\rm F}^2.
\end{eqnarray}
We use the  variational trial function
\begin{eqnarray}\label{ansatz}
\phi_{\rm B,F}(x)=\frac{1}{\pi^{1/4}}\sqrt{\frac{{\cal N}_{\rm
B,F}}{w_{\rm B,F}}} \exp\left(-\frac{x^2} {2w^2_{\rm B,F}}\right) ,
\end{eqnarray}
where the parameters $w_{\rm B,F}$ are the widths of the Bose and
Fermi localized states, respectively, with normalization ${\cal N_{\rm
B,F}}=\int_{-\infty}^{\infty}\phi ^2_{\rm B,F}(x) dx$.
The trial function (\ref{ansatz}) and
potential (\ref{pot})  lead to the  the effective Lagrangian
\begin{eqnarray}\label{lag}
L&=&\int_{-\infty}^{\infty}\mathscr{L}dx -\mu_{\rm
B}N_{\rm B}-\frac{1}{2}\mu_{\rm F}N_{\rm F}  \nonumber  \\
 &=&\mu_{\rm B}N_{\rm B}({\cal N}_{\rm B}-1)
 +\frac{1}{2}\mu_{\rm F}N_{\rm F}({\cal N}_{\rm F}-1)
 -\frac{N_{\rm B}{\cal N}_{\rm B}}{4w_{\rm B}^2}\nonumber\\
&&-\frac{N_{\rm F}{\cal N}_{\rm F}}{16w_{\rm F}^2}-\frac{g_{\rm
B}N_{\rm B}{\cal N}_{\rm B}^2}{2\sqrt{2\pi}w_{\rm
B}}-\left(\frac{3}{5}\right)^{3/2} \frac{g_{\rm F}N_{\rm F}{\cal
N}_{\rm F}^{5/3}}{\pi^{1/3} w_{\rm F}^{2/3}} \nonumber\\
&&-\frac{g_{\rm BF}N_{\rm B}N_{\rm F}{\cal N}_{\rm B}{\cal N}_{\rm
F}}{\sqrt{\pi}\sqrt{w_{\rm B}^2+w_{\rm F}^2}}+L_{\rm B}+L_{\rm F}, \\
&& L_{\rm B, F}=-\frac{N_{\rm B, F}{\cal N}_{\rm B, F}}{2}\sum_{l=1}^2
A_l \left(1-e^{-k_l^2w_{\rm B, F}^2}\right).
\end{eqnarray}
Euler-Lagrange equations, $\partial L/\partial \mu_{\rm B,
F} =0$ yield: ${\cal N}_{\rm B}={\cal N}_{\rm
F}=1$.
The remaining Euler-Lagrange
equations $\partial L/\partial w_{\rm B, F}
= 0$ yield, respectively,
\begin{eqnarray}
1+\frac{g_{\rm B}w_{\rm B}}{\sqrt{2\pi}}-2W_{\rm B}+2N_{\rm
F}G_{\rm B}=0, \label{width1}\\
1+\frac{16}{3}\left(\frac{3}{5}\right)^{3/2}\frac{g_{\rm F}w_{\rm
F}^{4/3}}{\pi^{1/3}}-8W_{\rm F}+8N_{\rm B}G_{\rm F}=0,
\label{width2}  \\
W_{\rm B,F}\equiv w_{\rm B,F}^4\sum_{l=1}^2A_lk_l^2e^{-k_l^2w_{\rm
B,F}^2}, \label{WBF}\\
G_{\rm B,F}\equiv \frac{g_{\rm BF}w_{\rm B,F}^4}{\sqrt{\pi}\left(w_{\rm
B}^2+w_{\rm F}^2\right)^{3/2}}, \label{GBF}
\end{eqnarray}
where we have set ${\cal N}_B={\cal N}_F=1. $
Equations (\ref{width1}) and (\ref{width2}) give the spatial widths
of the localized states.

\begin{figure}
\begin{center}
\includegraphics[width=\linewidth,clip]{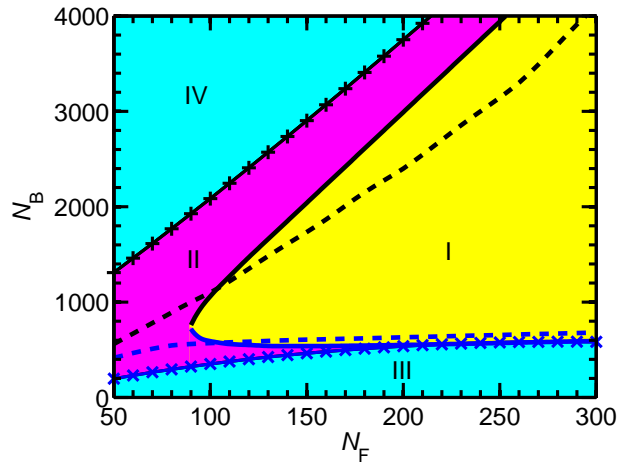}
\end{center}

\caption{(Color online) The phase diagram of Bose and Fermi
atom numbers  showing the regions of localization without
external trap (region I) and with bichromatic OL (\ref{pot})
(regions I and II)
from a solution of
the variational equations
(\ref{width1}) and (\ref{width2}).
In regions III and IV the variational equations have no solution.
Single-peaked localization in the  bichromatic OL
is possible  from a numerical solution of Eqs.
(\ref{model1}) and (\ref{model2}) in the region bounded by the dashed lines.
} \label{fig1}
\end{figure}

To study how widths $w_{\rm B,F}$ evolve with the particle
numbers $N_{\rm B,F}$ we  solve Eqs. (\ref{width1}) and
(\ref{width2}) and show  the phase diagram of
 the number of bosons and fermions in Fig. \ref{fig1}
illustrating  the regions where the
localization can exist with or without trapping potential
(\ref{pot}). The region I  bounded
by the solid line
corresponds to $V(x)=0$ where Eqs. (\ref{width1})
and (\ref{width2}) have  finite real solutions for both widths
$w_{\rm B, F}$. In this region, symbiotic localized states supported
by inter-species Bose-Fermi attraction can be
produced in spite of intra-species Bose and Fermi
repulsion. Generally, a self-repulsive Bose or Fermi super-fluid
cannot support a localized state by itself, however, a sufficiently
strong inter-species attraction can induce a net effective attraction
responsible for the formation of symbiotic
states in Bose-Fermi system \cite{PLA-346-179}.

With bichromatic OL   (\ref{pot}),  Eqs. (\ref{width1}) and
(\ref{width2}) yield  real solutions
for both widths $w_{\rm B, F}$ in regions I and II
bounded by the lines with crosses. In
regions III  and IV, however,  finite real
solutions of Eqs.  (\ref{width1}) and
(\ref{width2})
do not exist.
In region II, the fact that the widths
$w_{\rm B, F}$ are finite and real implies that
the localized
states are created by the quasi-periodic
bichromatic OL (\ref{pot}).
The parameter space
in Fig. \ref{fig1}
where localization can exist is enlarged from region I
to region II because
of the bichromatic OL. The phase diagram   also shows that
the effect of the bichromatic
OL is small when $N_{\rm
B}$ is small and $N_{\rm F}>200$.
For example, the region II between regions I and III
is very narrow. This implies that,
compared to  the  bichromatic OL,
the inter-species interaction is dominant in
this region, because   a larger $N_{\rm F}$ and small $N_B$
induce a large enough effective inter-species
attraction to localize the Bose-Fermi mixture.
When  $N_{\rm B}$ is larger, however, the effect of the bichromatic OL is
more important on localization as
 indicated by a wide region II between regions I
and IV.

The present discussion based on a Gaussian ansatz for the wave function
has its limitation in the presence of the bichromatic OL where it is
possible to have multi-peak density for the components not taken care of by
the simple Gaussian ansatz. Single-peaked Gaussian-type localization is
possible in regions I and II. However, in the presence of bichromatic OL,
localized states can exist
beyond regions I and II  into regions III and IV, where at least one of the
components has a multi-peak density distribution along the OL. Such states
are not obtainable from  a Gaussian variational analysis and will be
studied in Sec. \ref{III} using the full numerical solution of
Eqs. (\ref{model1}) and (\ref{model2}).

\section{Numerical results}
\label{III}

We perform the numerical integration of coupled GPEs (\ref {model1})
and (\ref {model2}) employing real- and imaginary-time propagation
using the split-step Fourier spectral method with space step $0.04$,
time step $0.001$.
The time evolution is continued till convergence.
We also checked the accuracy of the results by varying the space and
time steps and the total number of space and time steps. Although
the imaginary-time propagation method could find some of the
localized states, the stability of these states were confirmed
through the real-time propagation method. For studying the dynamics
we used real-time propagation corresponding to the solution of the
full time-dependent Eqs. (\ref{model1}) and (\ref{model2}).

\begin{figure}
\begin{center}
\includegraphics[width=.49\linewidth,clip]{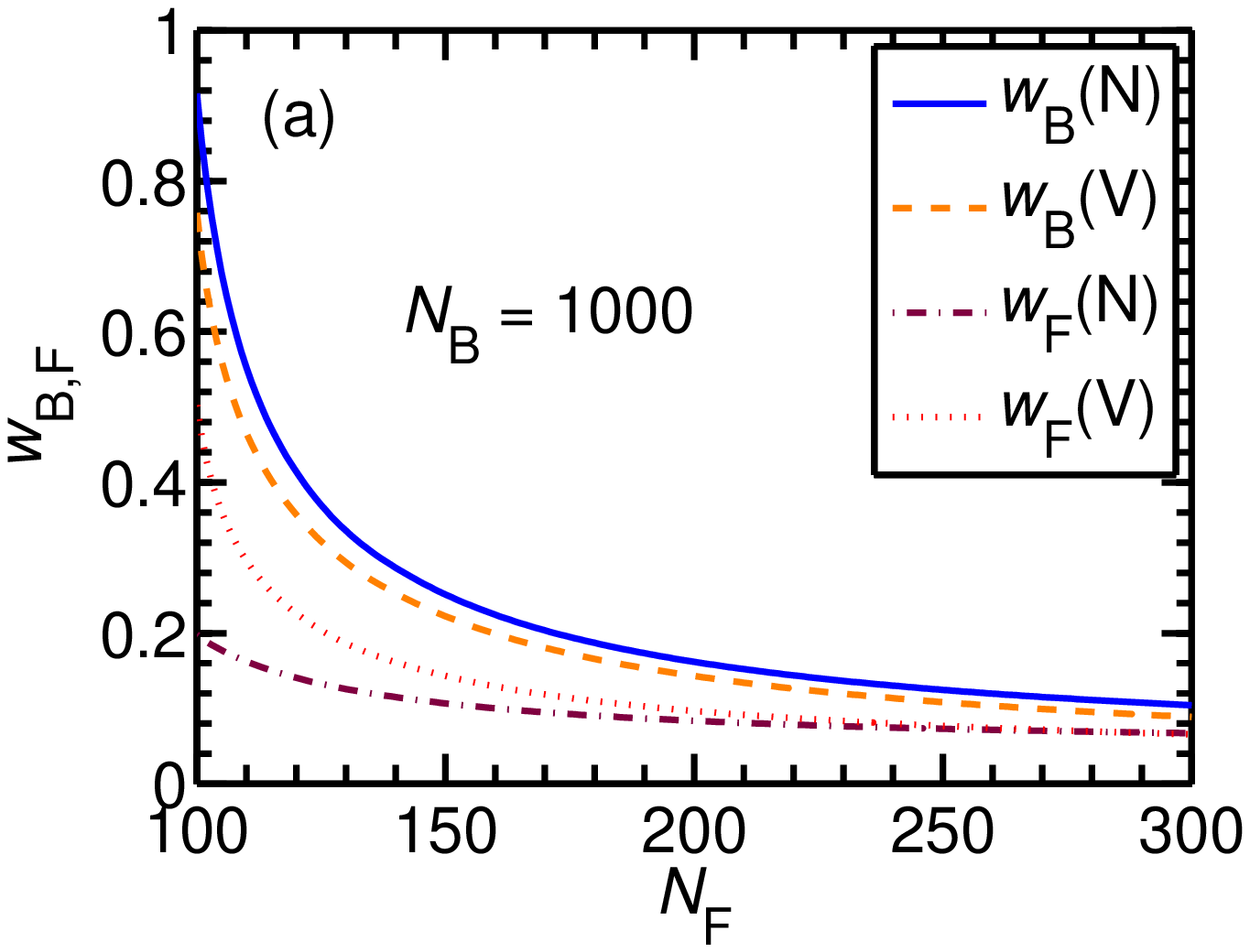}
\includegraphics[width=.49\linewidth,clip]{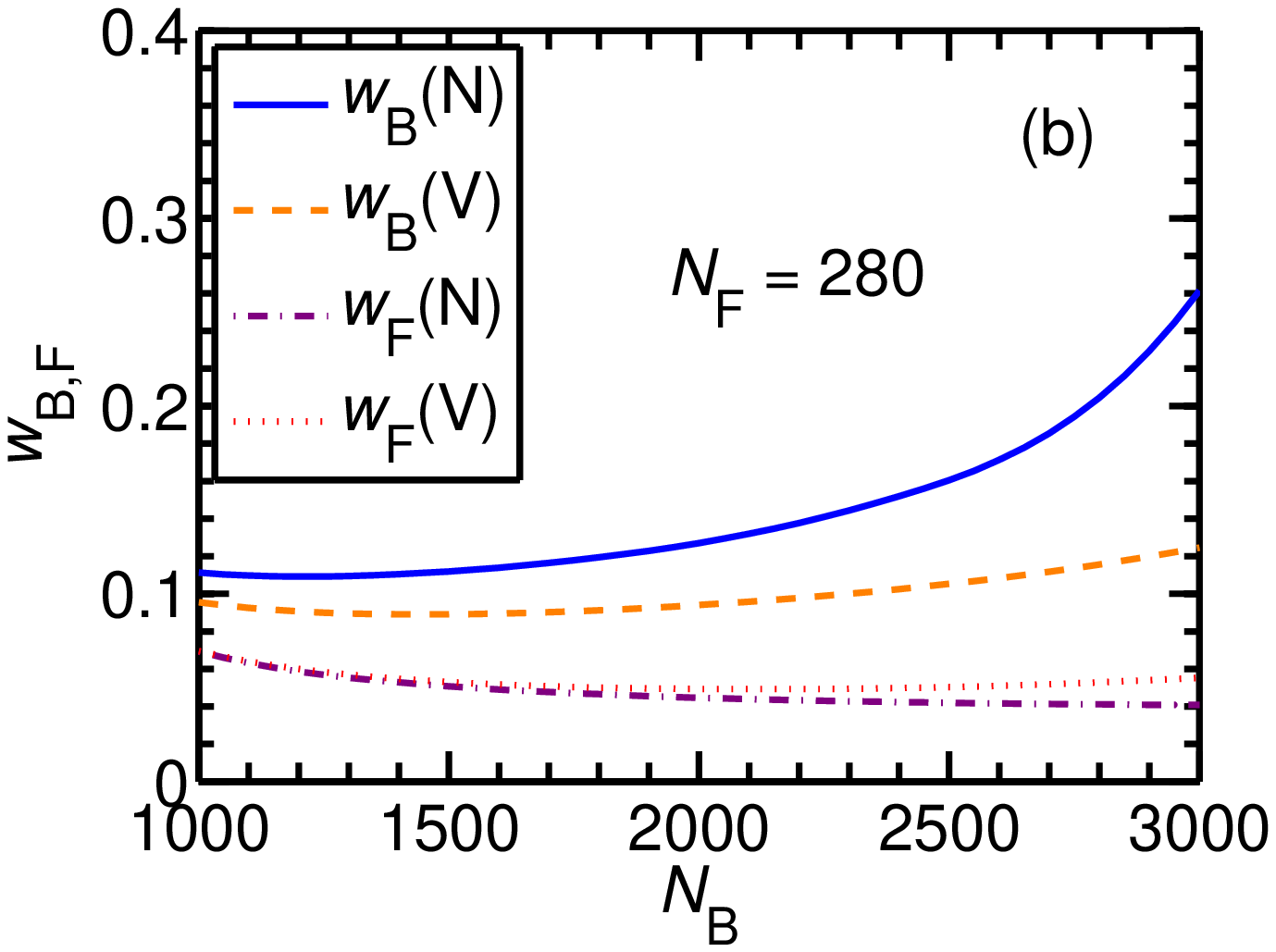}
\end{center}

\caption{(Color online) Dimensionless numerical (N) and
variational (V) widths $w_{\rm B,F}$ of the
Bose and Fermi components of the
stationary symbiotic localized states with $V(x)=0$
vs. (a) $N_{\rm F}$ for
$N_{\rm B}=1000$, and vs. (b) $N_{\rm B}$ for $N_{\rm F}=280$. }
\label{fig2}
\end{figure}

We first confirm the existence of the symbiotic localized states. By
numerical integration of GPEs (\ref{model1}) and (\ref {model2})
with $V(x)=0$, we find that the symbiotic localized states of the BF
systems can exist in Region I of Fig. \ref{fig1}. In order to
investigate the effects of the number of atoms  of the two components
on the localized states, the typical numerical (N) and variational
(V) widths of the atom density profiles of the stationary symbiotic
localized states are exhibited in Fig. \ref{fig2}.
The numerical widths are calculated via $w^2_{\rm
B,F}=2\int_{-\infty}^{+\infty}x^2|u_{\rm B,F}|^2 dx$. We find that the numerical
results are in good agreement with the variational results in the central
part of region I. Near the edges (the solid line in Fig. \ref{fig1}), the
difference between the numerical and variational widths is larger
because of the deformation of the wave functions, while the atom density
distributions deviate from the single-peak
Gaussian shape and assume a multi-peak structure.
  When $N_{\rm B,F}$ are smaller, we find that the deformation of the
fermionic component is larger than that of the bosonic component consistent
with
Fig. \ref{fig2} (a). On the contrary,
the deformation of the bosonic component is larger than that of the
fermionic component when $N_{\rm B,F}$ are larger consistent with Fig.
\ref{fig2} (b).

\begin{figure}
\begin{center}
\includegraphics[width=.49\linewidth,clip]{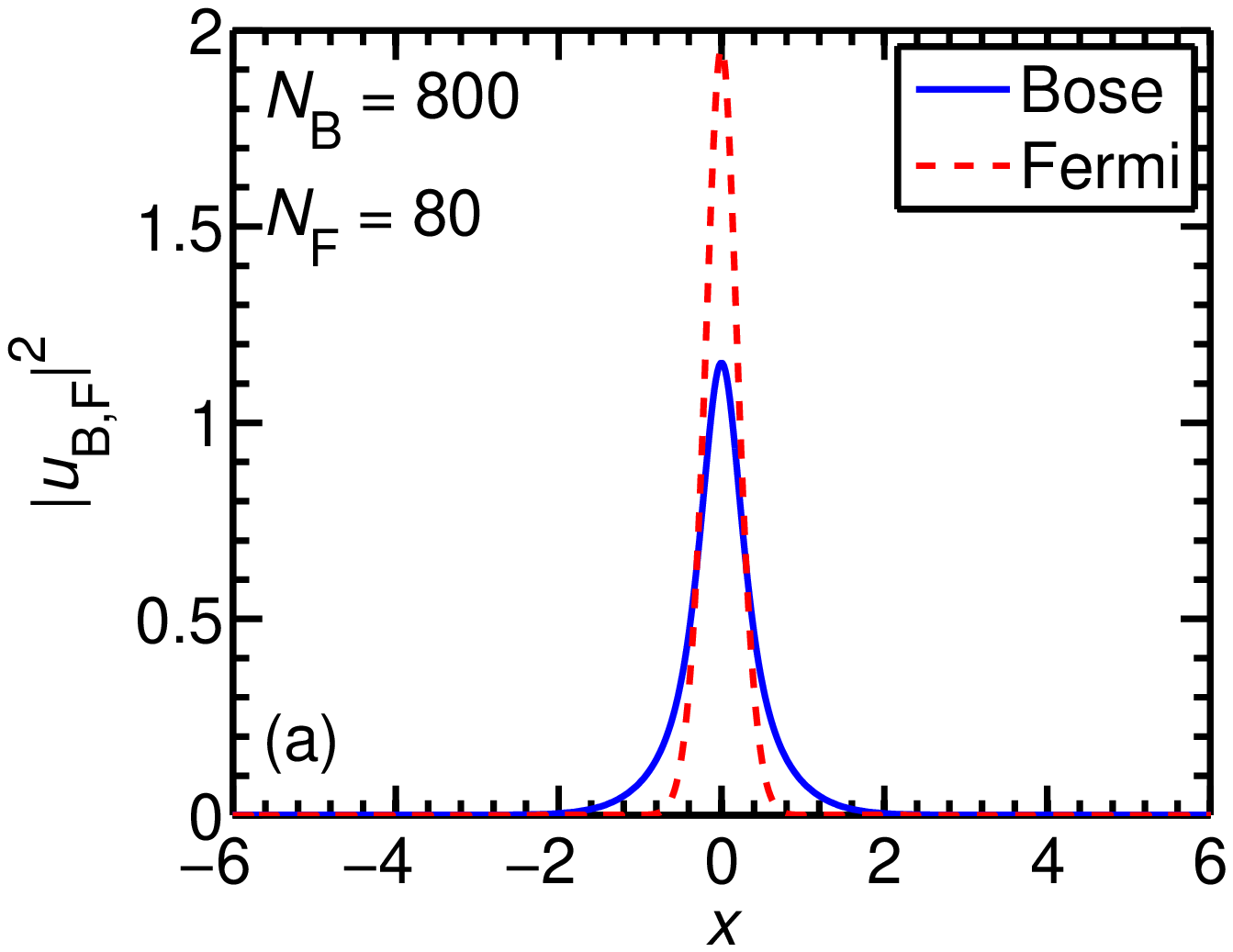}
\includegraphics[width=.49\linewidth,clip]{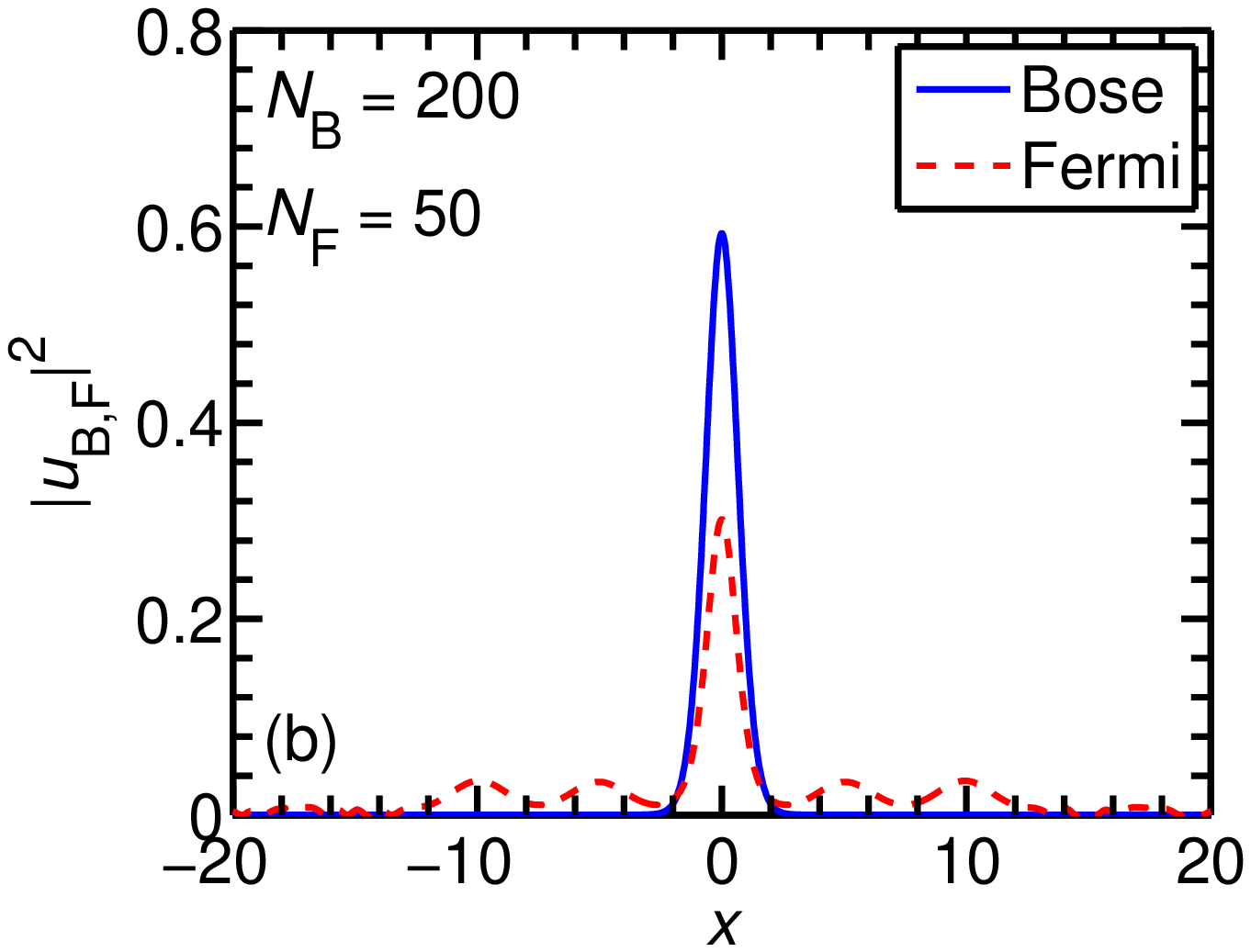}
\includegraphics[width=.49\linewidth,clip]{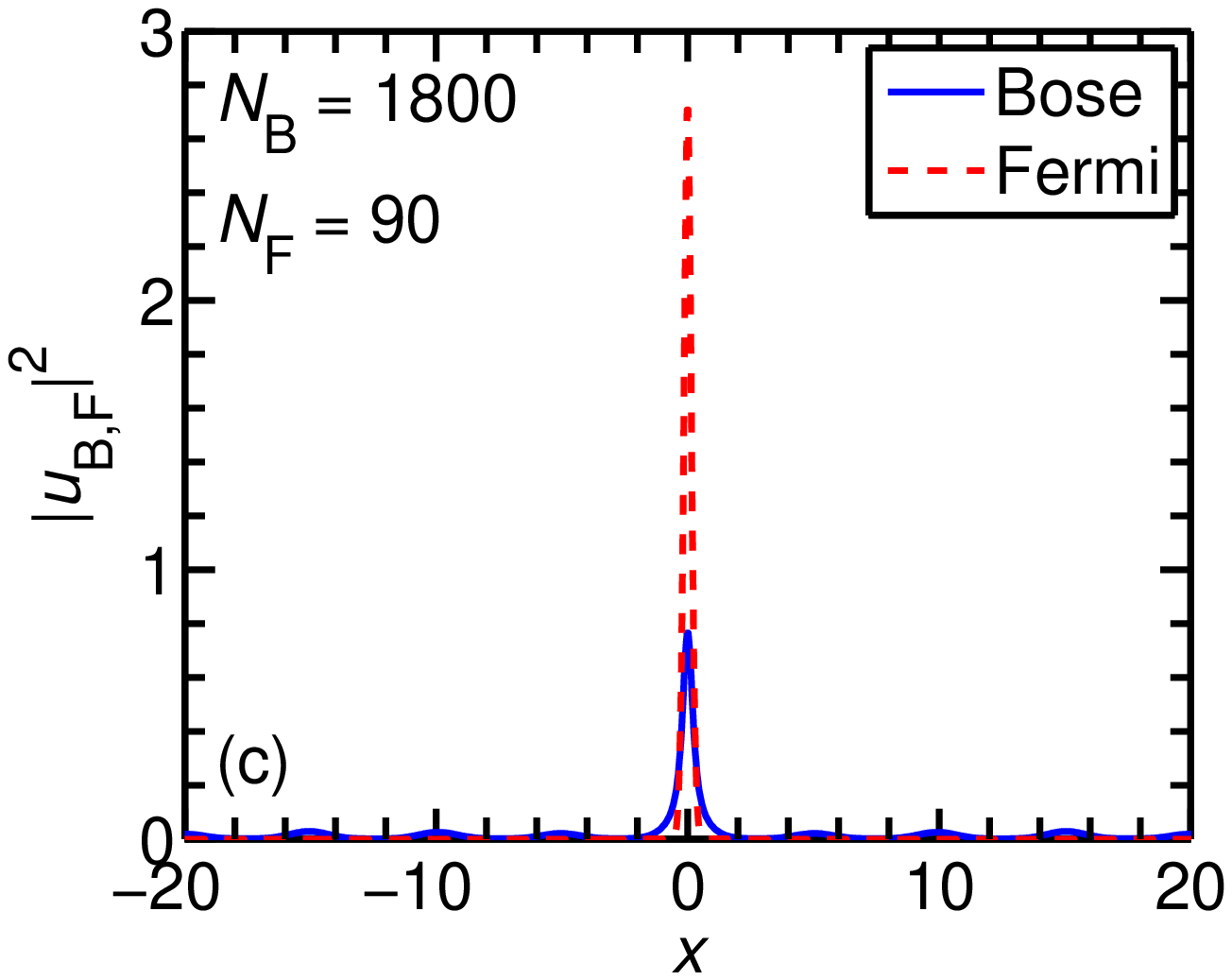}
\includegraphics[width=.49\linewidth,clip]{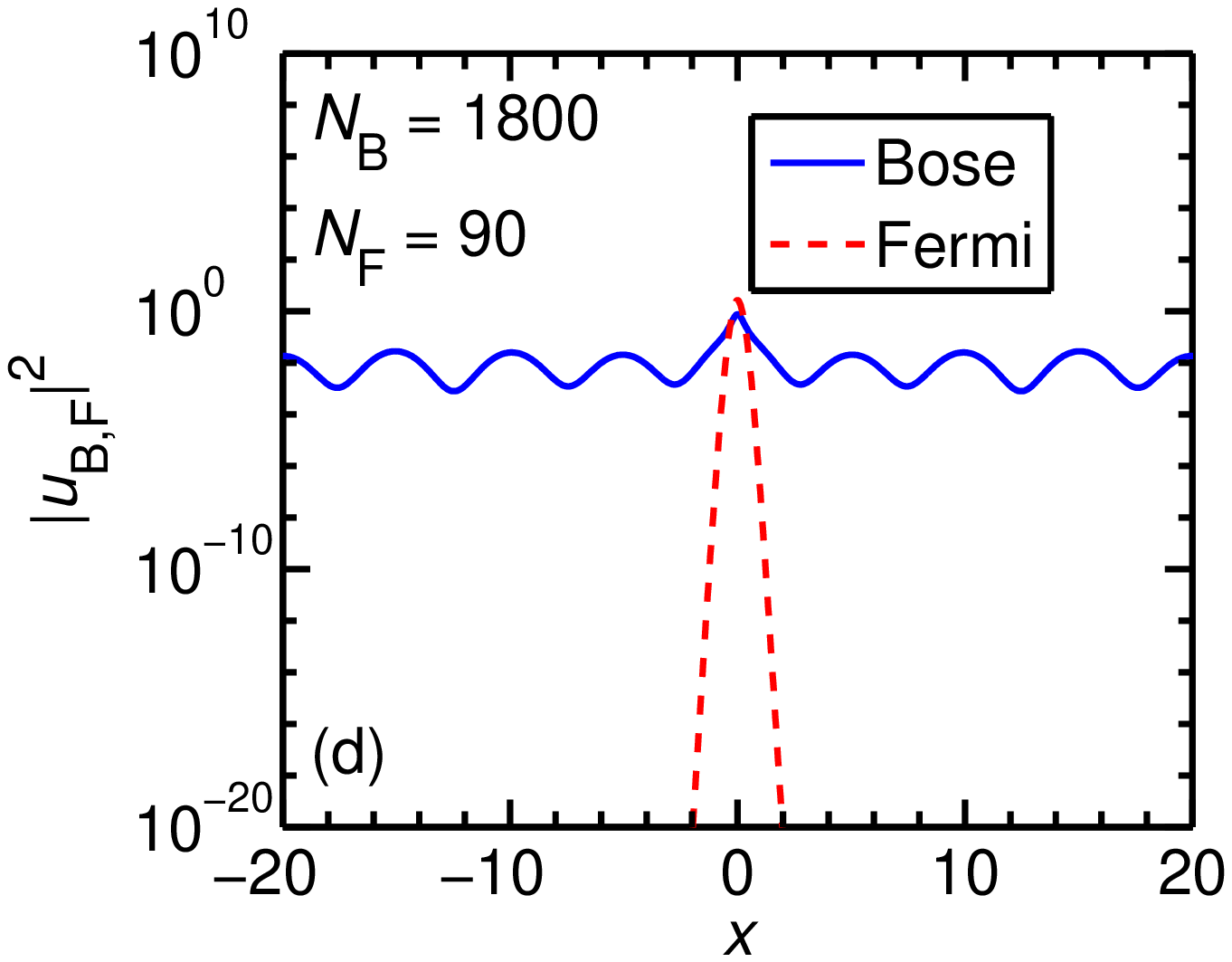}
\end{center}

\caption{(Color online) Numerical  densities $|u_{\rm
B,F}|^2$ of the coupled Bose-Fermi localized states with the
bichromatic OL  (\ref{pot}) vs. dimensionless position
$x$ for (a) $N_{\rm B}=800,
N_{\rm F}=80$, (b) $N_{\rm B}=200, N_{\rm F}=50$, (c) $N_{\rm
B}=1800, N_{\rm F}=90$, (d) $N_{\rm B}=1800, N_{\rm F}=90$ in
logarithmic scale. The solid line corresponds to  bosons and the
dashed line  to  fermions.}
\label{fig3}
\end{figure}

A careful analysis of the density distribution of the localized
Bose-Fermi system reveals some interesting features.
In fact,
an attractive inter-species interaction may induce a
spatially localized effective attraction
 \cite{JPB-42-205005} which can reduce
the width of the atom density profiles, and even develop a sharp
peak on the top of the bosonic and fermionic atom density envelopes
\cite{PRA-71-033609} as we shall see in the following,
viz. Fig. \ref{fig3}. Once  a peak appears in the density of the
first   species (Bose or Fermi), this induces a highly localized
effective interaction in the second species (Fermi or Bose)
resulting in a highly localized structure in the second species.
The small width of a species, say Bose,
may, however,  emerge in two ways: (i) a large overall attractive inter-species
interaction   $g_{\rm BF}N_{\rm F}|u_{\rm F}|^2u_{\rm B}$
due to a large number of the second species, $N_{\rm F}$, or (ii) a
highly localized inter-species interaction $g_{\rm BF}N_{\rm
F}|u_{\rm F}|^2u_{\rm B}$ due to  a narrow localized state of the
second species, Fermi.

Figure \ref{fig2} (a) indicates that $w_{\rm B,F}$ decreases
monotonically as $N_{\rm F}$ increases. This is because a larger
$N_{\rm F}$ induces a lager effective attraction among bosons
\cite{PLA-346-179}, which reduces the  Bose width $w_{\rm B}$
due to possibility (i) above. Simultaneously, the narrower Bose
localization causes a stronger attractive point-like effective
potential for the Fermi component, which reduces the Fermi
width $w_{\rm F}$ due to possibility (ii). Figure
\ref{fig2} (b) shows that, as $N_{\rm B}$ increases, $w_{\rm F}$
decreases and $w_{\rm B}$ increases. With the increase of $N_{\rm
B}$ for a fixed $N_{\rm F}$, the Bose species becomes more repulsive
due to an increase of intra-species repulsion for a roughly fixed
inter-species attraction thus increasing the width of the Bose
species. However, the increase of $N_{\rm B}$ for a fixed $N_{\rm
F}$ increases the inter-species attraction on the Fermi species thus
reducing the Fermi width.


Next, we study the  Bose-Fermi localized states
with quasi-periodic OL
(\ref{pot}) in
region  II  of Fig. \ref{fig1}.
Typical atom density profiles of  stationary Bose and Fermi
localized states are shown in Figs. \ref{fig3} (a) $-$ (d). We find that,
because of the bichromatic OL, the localized
  Bose or Fermi states may exist beyond   region I.
For example, Fig. \ref{fig3} (a) shows both Bose and Fermi components
in single-peaked configuration  for parameters corresponding to
region II. The numerical simulation indicate, however, that both
components can be  single-peaked only in the region between the
dashed lines in Fig. \ref{fig1}. We name
this region the single-fragment region. In the region below the
single-fragment region, (viz., the region with smaller $N_{\rm B}$,)
the atom density profile of the Bose component is  single peaked, but
the Fermi component shows a symmetrical multi-peak structure with a
pronounced peak at center, as shown in Fig. \ref{fig3} (b).
However, in the region above the single-fragment region, (viz., the
region with larger $N_{\rm B}$,) the density profile of the Fermi
component is single-peaked and  that of the Bose component is
multi-peaked as shown in  Figs. \ref{fig3}(c) and (d). To show this
  clearly, Fig. \ref{fig3}(d) is plotted in a logarithmic scale
which shows that the Bose density profile   develops
undulating tails occupying many OL sites.
When $N_{\rm B}$  is larger, as in Figs. \ref{fig3} (c) and (d),  the Bose
system becomes more repulsive and hence spreads to many OL sites and thus
develops undulating tails. However, a large  $N_{\rm B}$ also introduces
a strong inter-species attraction on the Fermi component, thus confining
it in a small region. This narrow Fermi peak creates a smaller narrow peak in the
Bose component due to inter-species attraction as seen in Fig. \ref{fig3} (c).
The role of Bose and Fermi is interchanged for a small $N_B$ as can be seen in
Fig. \ref{fig3} (b), where one has a single-peak Bose distribution on top of a
multi-peak Fermi distribution.  The situation for an intermediate value of
$N_B$ is shown in Fig. \ref{fig3} (a).

Now we consider the possibility of Anderson localization of the bosons
with exponential tail in density
in the interacting
Bose-Fermi mixture on bichromatic OL (\ref{pot})
\cite{andlocus1c,adhisala,NJP-11-033023,NAT-453-895,billy}.
There are domains around region II
of phase diagram shown in Fig. \ref{fig1} where the inter- and intra-species
interaction on the bosons approximately
``cancel" each other. Consequently, the bosons behave like quasi-free particles
(with effective weak repulsion)
  and the weak  quasi-periodic
bichromatic OL
is necessary for localizing the bosons.
In order to understand the novel phenomenon, we investigate the
localization of Bose and Fermi components in the
 bichromatic OL (\ref{pot}) away
from the domain of strong inter-species interactions, where
there is no localization in the absence of the bichromatic OL
potential, e.g., around region II of Fig. \ref{fig1}.
In that case,  we find that the wave function of the Bose component may
possess a pronounced exponential tail   demonstrating Anderson localization. However, we
could not find a similar exponential tail of fermion
density distribution in the whole phase diagram in Fig. \ref{fig1}.

\begin{figure}
\begin{center}
\includegraphics[width=.49\linewidth]{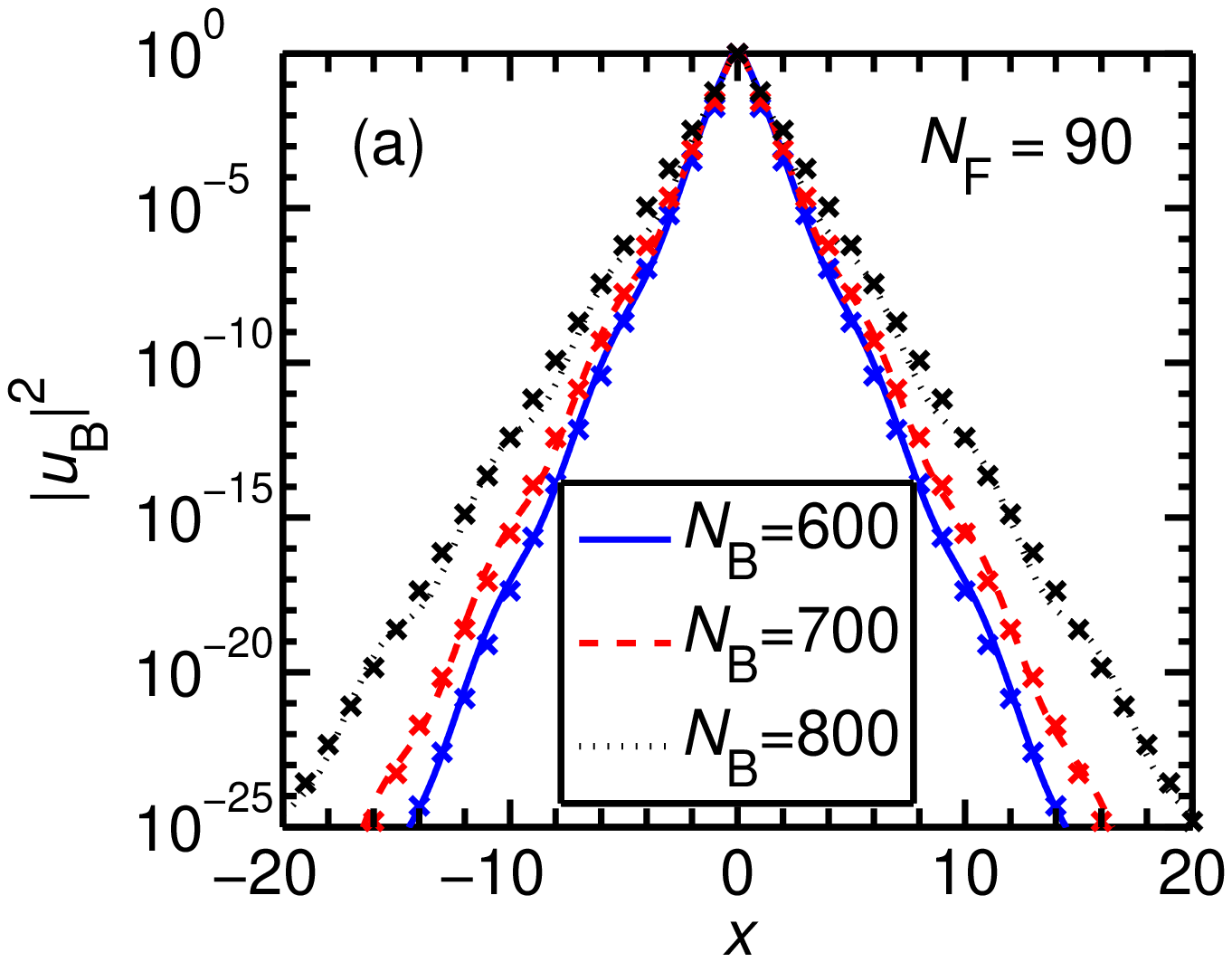}
\includegraphics[width=.49\linewidth]{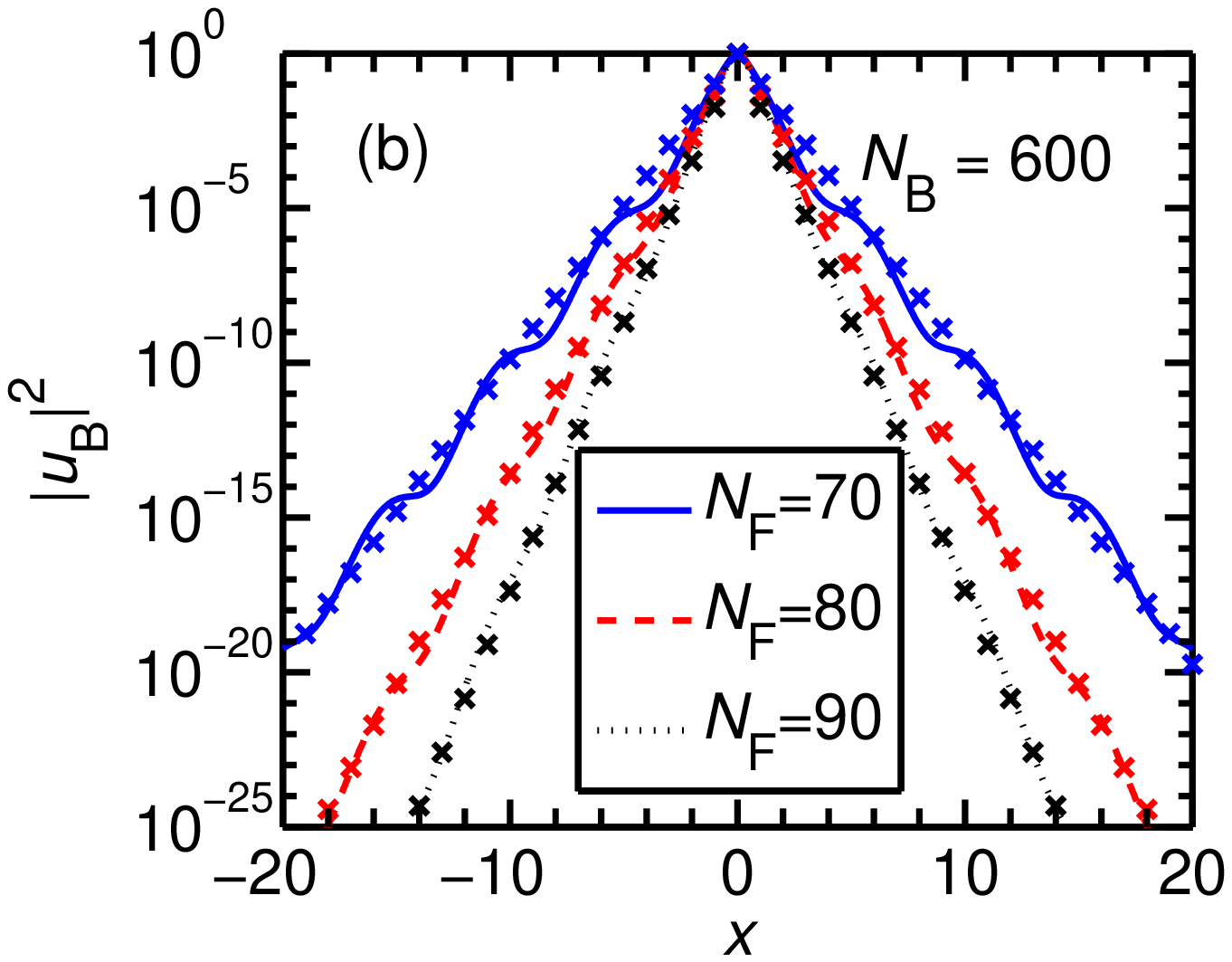}
\end{center}

\caption{(Color online)  Numerical density  $|u_{\rm
B}|^2$ of the Bose component in the Bose-Fermi mixture
on bichromatic OL
(\ref{pot}) vs. dimensionless $x$ in logarithmic scale for (a)
$N_{\rm F}=90$ and $N_{\rm B}=600$ (solid line), 700 (dashed line), 800
(dotted line), and for (b)
$N_{\rm B}=600$ and $N_{\rm F}=70$ (solid line), 80 (dashed line),
90 (dotted line). The
crosses are exponential fit to density tails   $\sim
\exp(-|x|/L_{\rm loc})$, where $L_{\rm loc}$ is the dimensionless
localization length. } \label{fig4}
\end{figure}

To study Anderson localization of the Bose component,   while
both Bose and Fermi components are single-peaked
under the action of OL trap  (\ref{pot}),  we plot in Fig.
\ref{fig4} the atom density  $|u_{\rm B}|^2$, in log
scale, of the stationary
Bose localized states. The central part of the atom
density is quasi-Gaussian, however,  with a long
exponential tail. We have also shown in Fig. \ref{fig4} the
exponential fitting to density tails with $\sim \exp(-|x|/L_{\rm
loc})$, where $L_{\rm loc}$ is
the localization length. (Note that our definition of localization length
\cite{andlocus1c}
differs from that of Refs. \cite{billy,adhisala} by a factor of 2.) The length $L_{\rm loc}$ depends
on the effective nonlinearity of the bosonic component and the
disorder of the quasi-periodic OL \cite{NJP-10-045019}.
Here, we fix the quasi-periodic OL (\ref{pot})
and study the effect of nonlinearity on
localization length. In Fig. \ref{fig4} (a), for
$N_{\rm F}=90$, $L_{\rm loc}$ is $=0.25$ for $N_{\rm
B}=600$, $=0.28$ for $N_{\rm B}=700$ and $=0.35$ for $N_{\rm B}=800$
and the localization length increases with $N_B$.
This is because that a larger $N_{\rm B}$, for a fixed $N_F$
implies a larger repulsive bosonic
nonlinearity, thus resulting in a larger
value of localization length.  Figure \ref{fig4} (b) indicates that,
for a fixed $N_B = 600$,
$L_{\rm loc}$ decreases with $N_{\rm F}$ with $L_{\rm loc}=0.44$ for
$N_{\rm F}=70$, $=0.32$ for $N_{\rm F}=80$ and $=0.25$ for $N_{\rm
F}=90$. A larger $N_{\rm F}$, for a fixed $N_B$, implies a larger
inter-species attraction on the bosonic component
 \cite{PLA-346-179}, thus
reducing the localization length.

\begin{figure}
\begin{center}
\includegraphics[width=.49\linewidth]{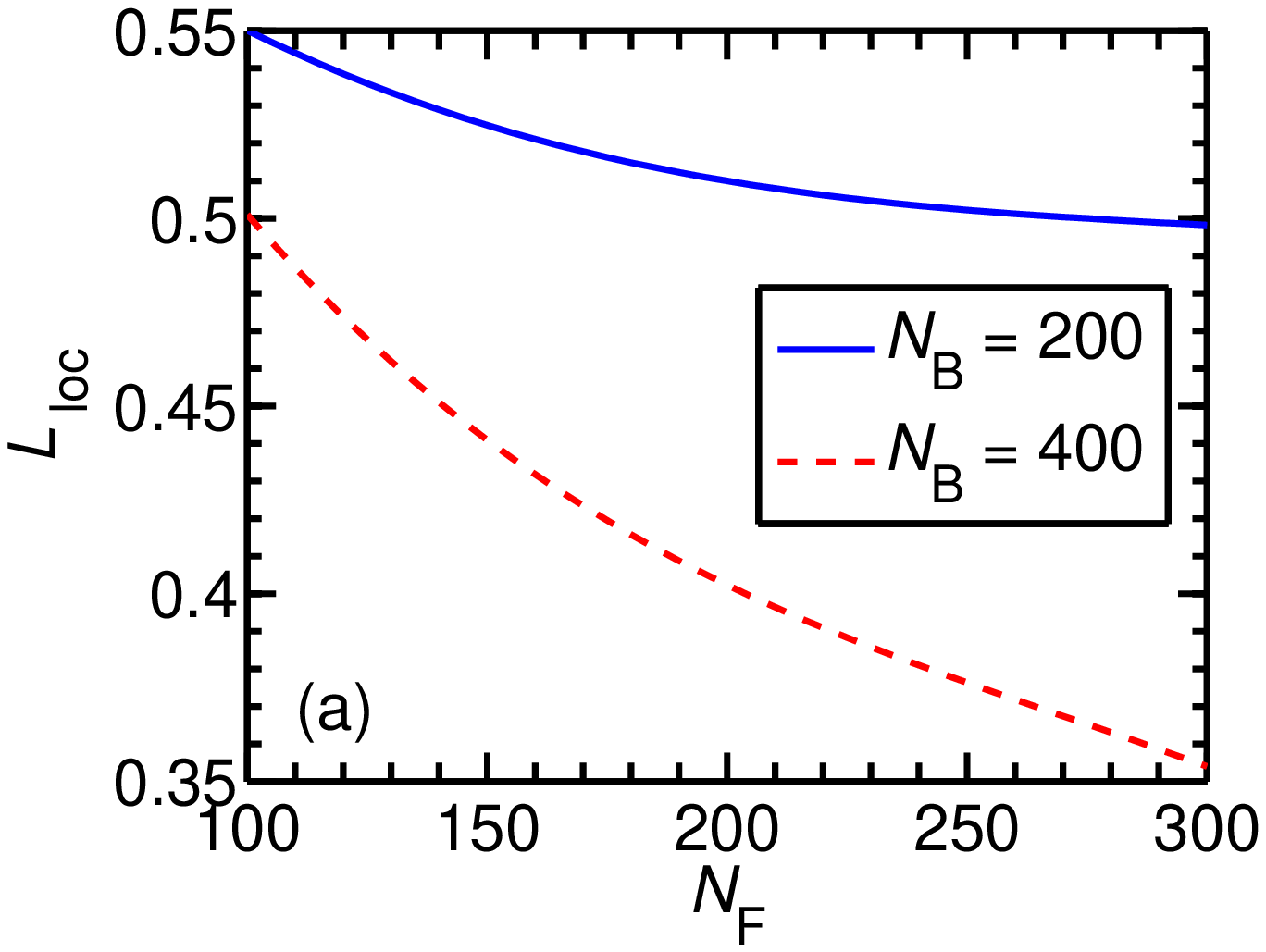}
\includegraphics[width=.49\linewidth]{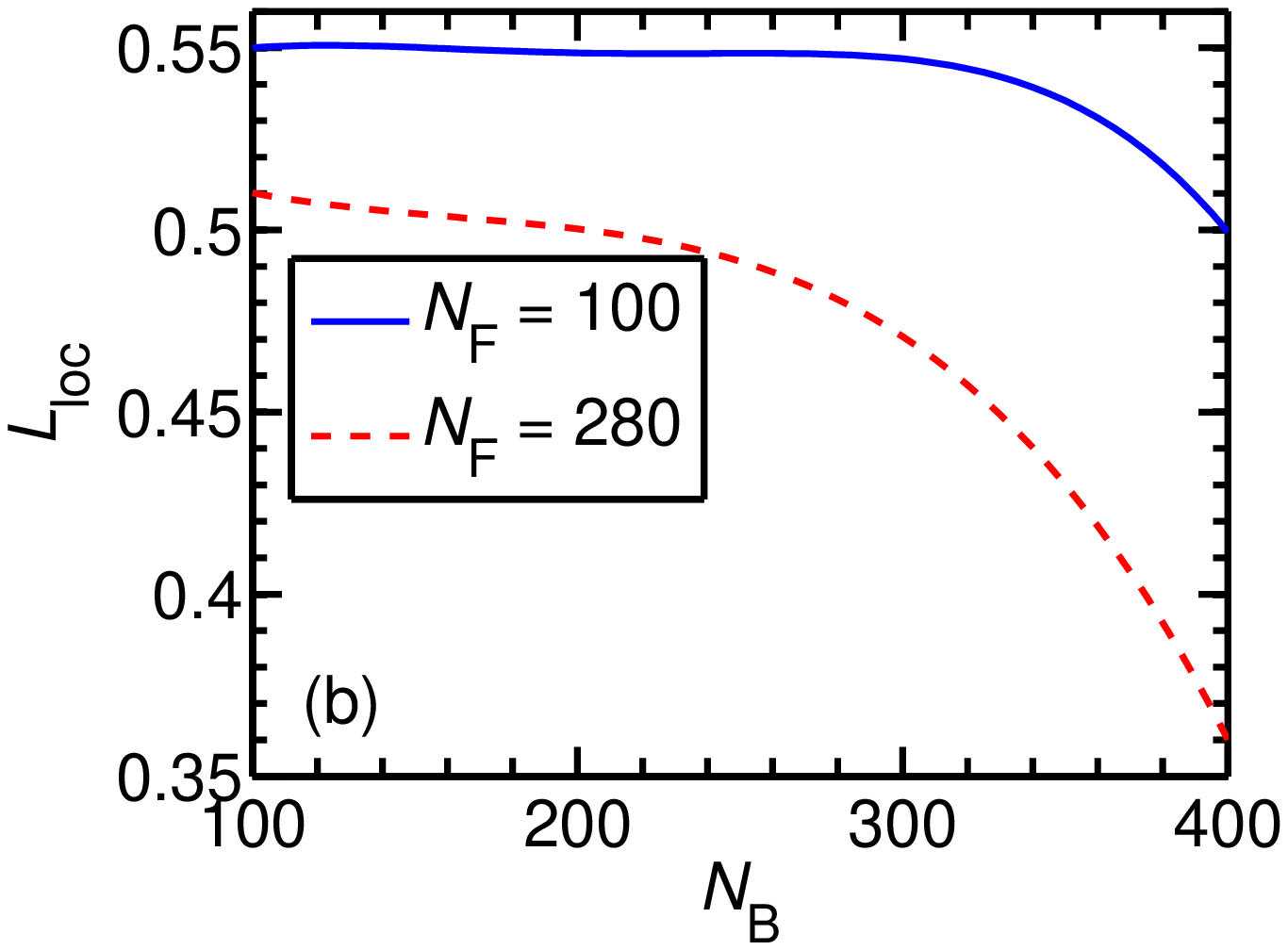}
\end{center}

\caption{(Color online) Dimensionless localization length
$L_{\rm loc}$ of Bose density $|u_{\rm B}|^2$ vs.
(a) $N_{\rm F}$ for $N_B=200,400$, and vs. (b)   $N_{\rm B}$ for
$N_F=100,280$, in the Bose-Fermi mixture on bichromatic OL
 (\ref{pot}).  }
\label{fig5}
\end{figure}

We also investigate numerically the Anderson localization of the
Bose component  (\ref{pot}) for small $N_B$. With OL   (3),
the Bose localization length $L_{\rm loc}$ versus $N_{\rm F}$
and $N_{\rm B}$ are illustrated in Figs. \ref{fig5} (a)
and (b), respectively. As shown in Fig.
\ref{fig5} (a), because of inter-species attraction, $L_{\rm loc}$
for a fixed $N_B$ monotonically decreases with the
increase of $N_{\rm F}$. For smaller
$N_B (=200)$, there is a saturation of  $L_{\rm loc}$ and Anderson localization
with exponential tail continues with the increase of $N_{\rm F}$.
For larger $N_B (=400)$, $L_{\rm loc}$ is reduced rapidly
with the increase of $N_F$ until the Anderson localization with exponential tail
is destroyed and one has strong localization without exponential tail.
In Fig. \ref{fig5} (b), we show the variation of $L_{\rm loc}$ with $N_B$
for $N_F=100$ and 280. From the phase plot in Fig. \ref{fig1} we find that
in both cases Anderson localization is  rapidly destroyed as we enter the
region I of Fig. \ref{fig1}
of permanent symbiotic trapping with the increase of $N_B$
and the localization length is reduced rapidly.

\begin{figure}
\begin{center}
\includegraphics[width=.49\linewidth]{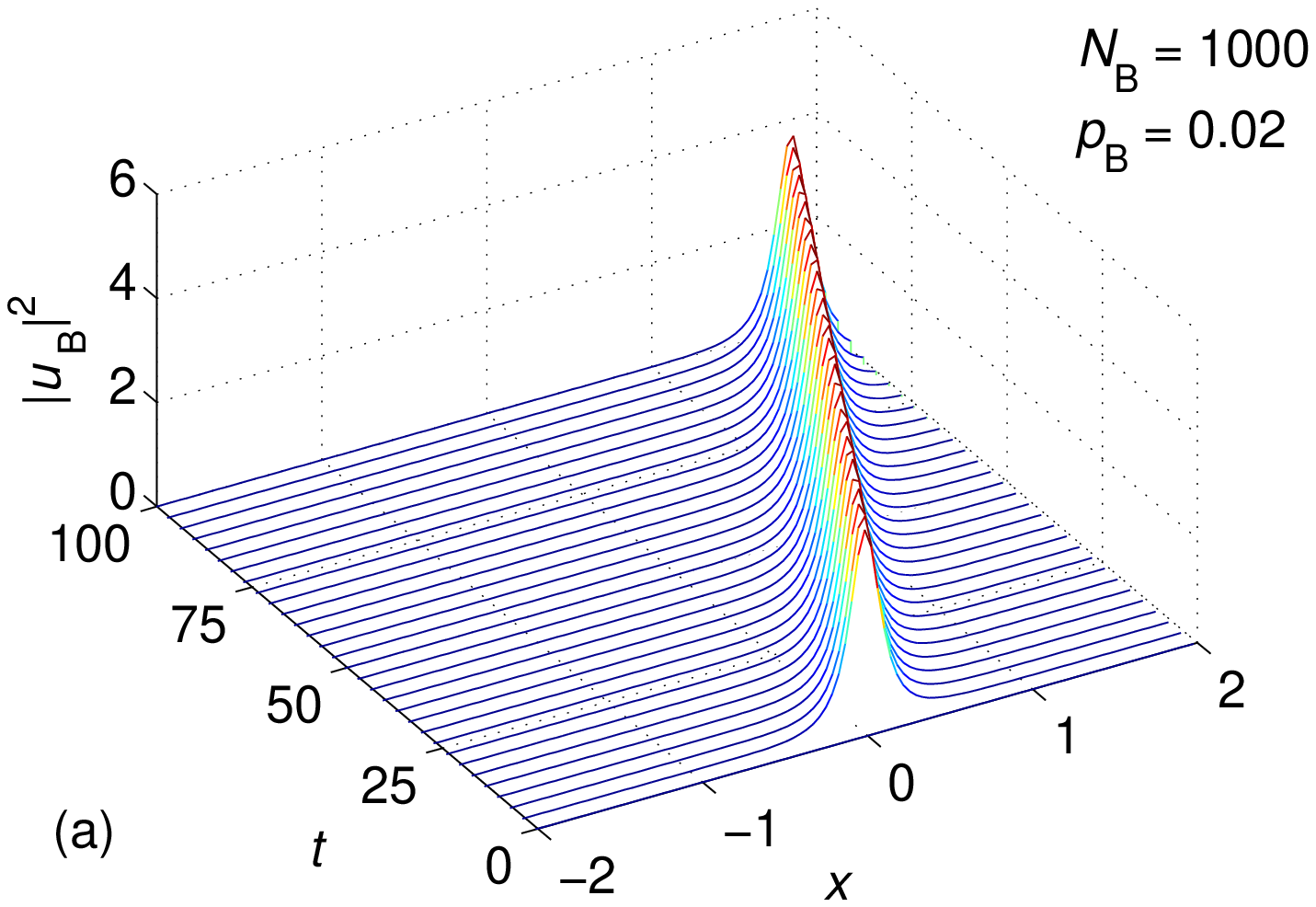}
\includegraphics[width=.49\linewidth]{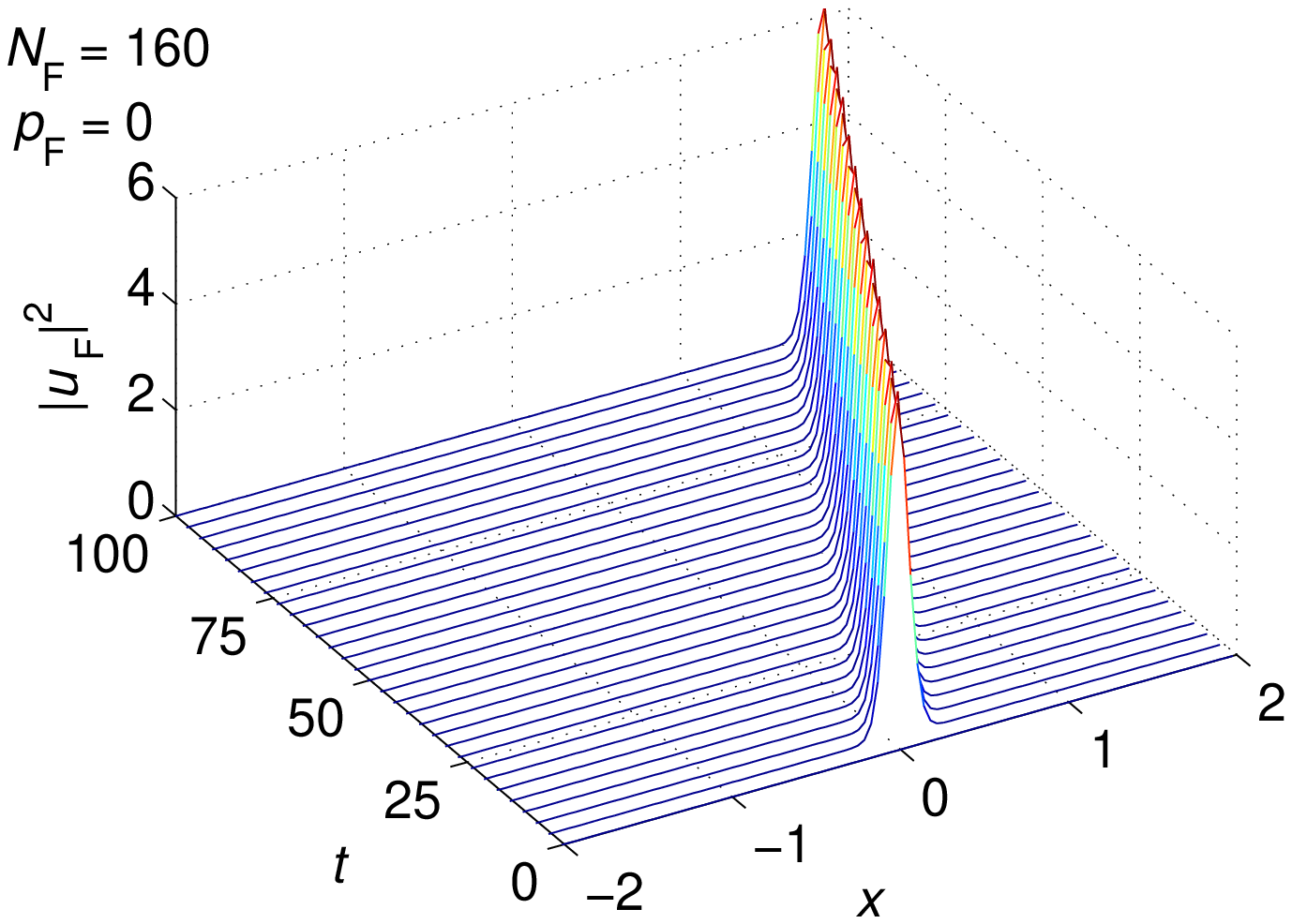}
\includegraphics[width=.49\linewidth]{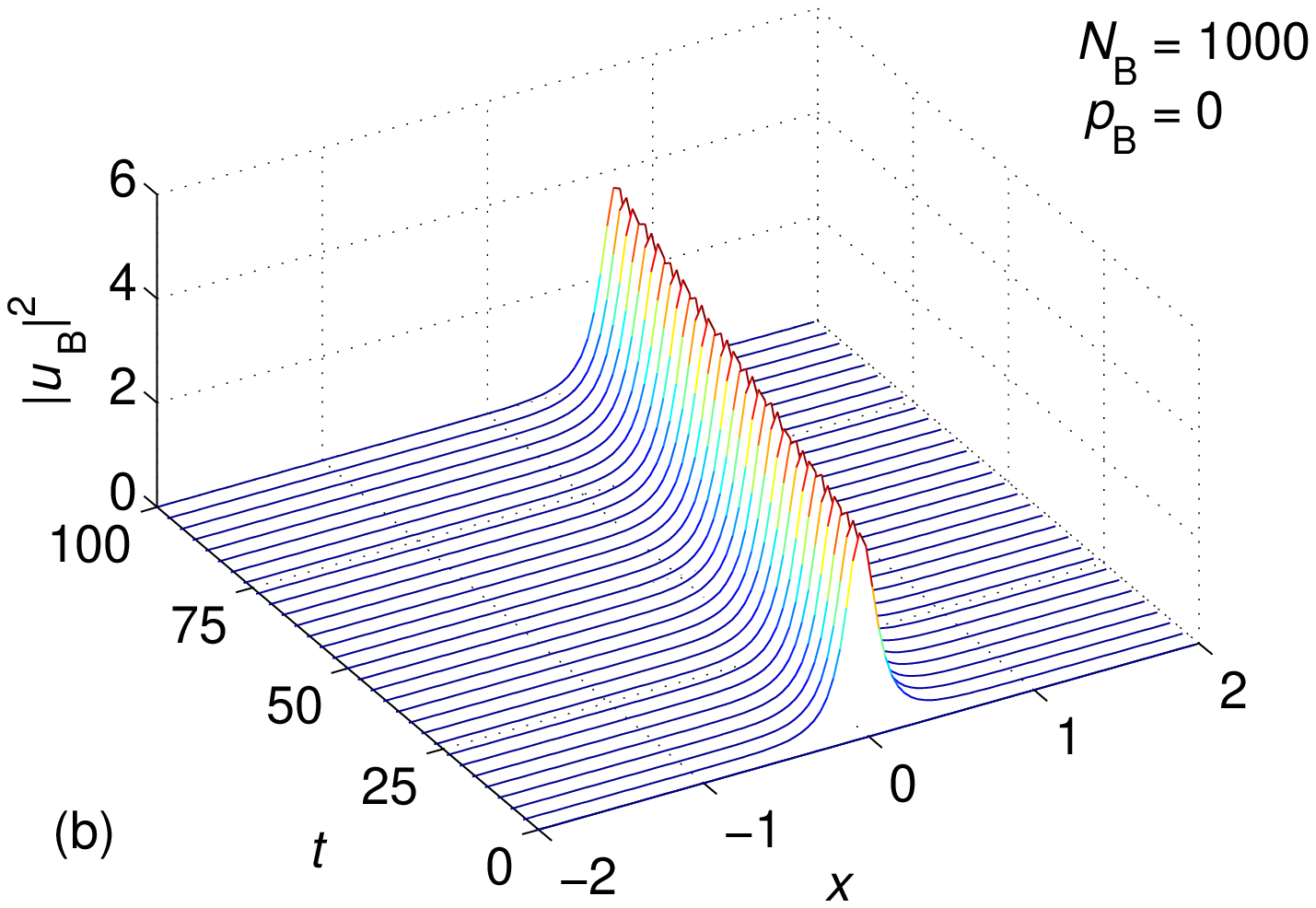}
\includegraphics[width=.49\linewidth]{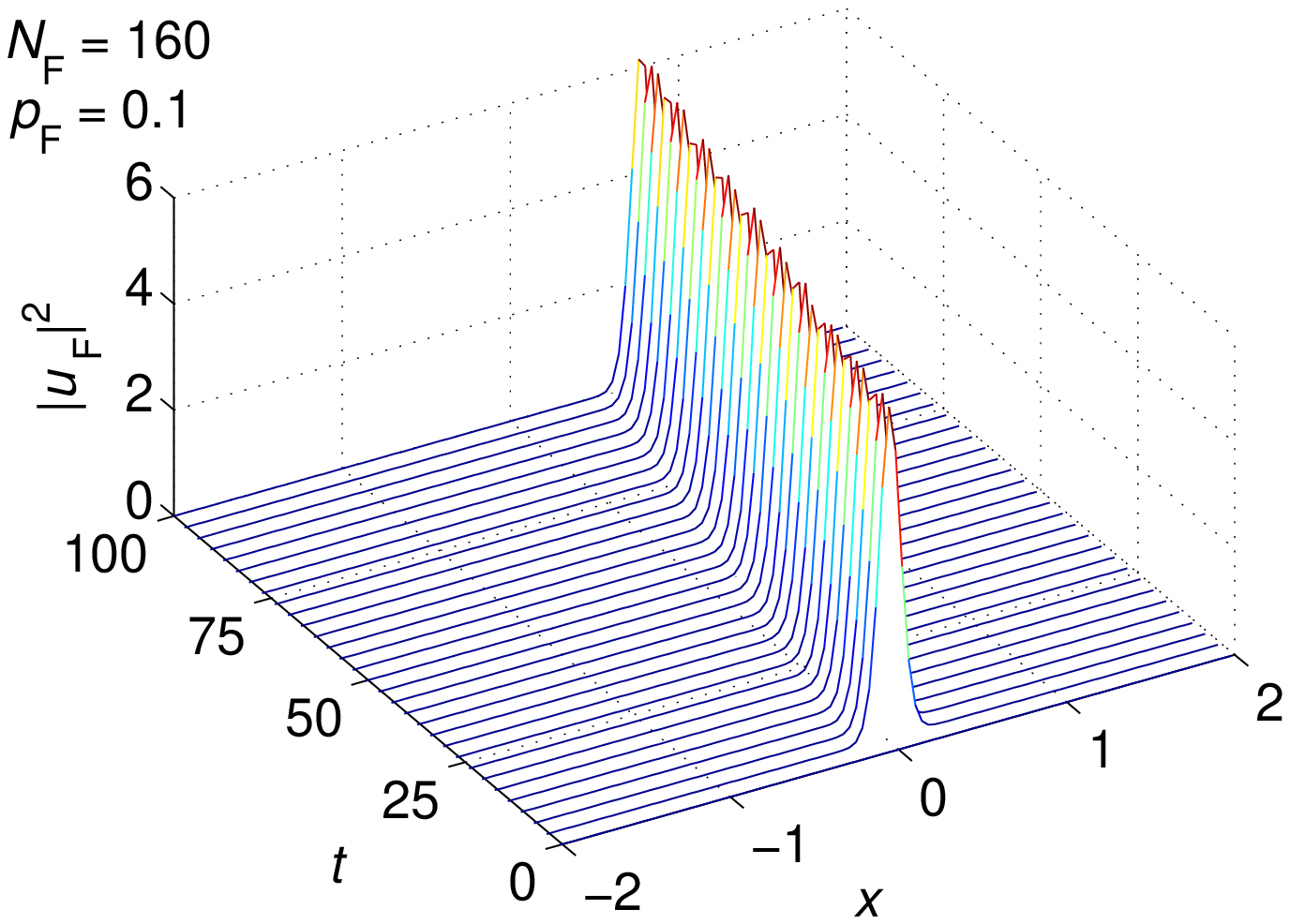}
\end{center}
\caption{(Color online) Numerical densities   $|u_{\rm
B,F}|^2$ vs. dimensionless $x$ and $t$ of the symbiotic Bose-Fermi localized
states  after introducing a momentum  in one of the components by
  transformation $u_{\rm B,F}\to u_{\rm B,F}\exp (ip_{\rm
B,F}x)$. The parameters are $N_{\rm
B}=1000, N_{\rm F}=160$ and
 (a)
$p_{\rm B}=0.02,p_{\rm F}=0$, (b) $p_{\rm B}=0,p_{\rm F}=0.1$.} \label{fig6}
\end{figure}

 Now we study numerically some
dynamics of the localized Bose-Fermi states and test their stability.
First, we consider a Bose-Fermi state in the region I of Fig. \ref{fig1}
corresponding to
permanent  symbiotic trapping.
To do this, first we create the stationary
localized Bose-Fermi mixture under appropriate conditions.
Successively, at
$t=0$, we suddenly introduce a phase $\exp(ip_{\rm B,F}x)$
in the wave function of one of the components
 to initiate a translational motion.
No momentum is given to the second component.
By introducing an initial translation
to the Bose or Fermi component, the evolution of the atom
density envelopes is presented in Fig. \ref{fig6} for $N_{\rm B}=1000,
N_{\rm F}=160$,  and for  (a)  $p_{\rm B}=0.02,
p_{\rm F}=0$, and for (b) $p_{\rm B}=0, p_{\rm F}=0.1$.
We find that the two components remain bound together and move
with the same constant velocity although the initial momentum is given
to only one component. The binding is caused by the
stronger inter-species attraction. We see in Fig. \ref{fig6}
that the symbiotic Bose-Fermi localized states remain
unchanged after perturbation,
which confirms the stability of these states.


\begin{figure}
\begin{center}
\includegraphics[width=\linewidth,clip]{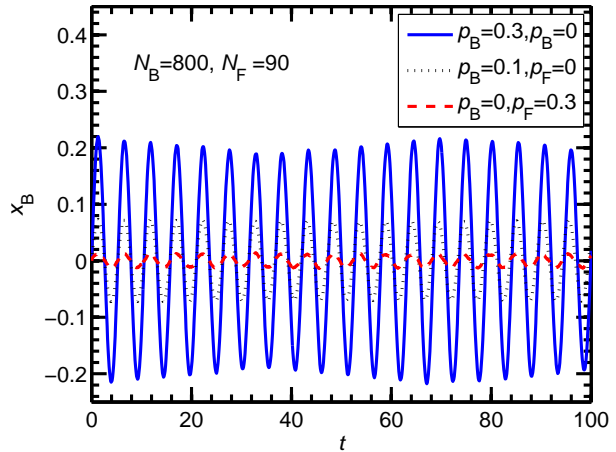}
\end{center}

\caption{(Color online) The dimensionless center of the
Bose state $x_{\rm B}$ vs. dimensionless time $t$
during the location oscillation of the Bose-Fermi mixture
on the bichromatic OL (\ref{pot})
initiated suddenly by introducing an
initial momentum $p_{\rm B,F}$ through  the transformation $u_{\rm
B,F}\to u_{\rm B,F}\exp (ip_{\rm B,F}x)$.    }
\label{fig7}
\end{figure}

Next we consider the dynamics of an Anderson localized Bose state in
the Bose-Fermi mixture on
the bichromatic OL (\ref{pot}) with $N_B=800$ and $N_F=90$
corresponding to one of the localized states of Fig.
\ref{fig4} (a). The dynamics is initiated by
introducing a phase $\exp(ip_{\rm B,F}x)$
in the wave function of one of the components
 to start a translational motion.
Because of the quasi-periodic OL (\ref{pot}), the
bound Bose-Fermi localized state oscillates together
periodically about the origin in one site of the OL
after introducing an initial momentum
into one of the components, as shown in Fig. \ref{fig7} which
shows only the movement of the center of the Bose component, e.g.,
$x_{\rm B}$ vs. $t$. Different initial momenta result in
oscillation with different amplitudes, see the solid line
($p_{\rm B}=0.3$), dashed line ($p_F=0.3$) and the dotted line
($p_{\rm B}=0.1$) in Fig. \ref{fig7}. Sustained oscillation of the
Bose (as well as the bound Fermi component not shown here)
component confirms the stability of the coupled Bose-Fermi
state.


\section{SUMMARY}
\label{IIII}

Here we studied symbiotic localization
in a cigar-shaped Bose-Fermi super-fluid with intra-species
repulsion and inter-species attraction.
The Fermi component is considered at unitarity. In the presence of
a quasi-periodic bichromatic OL, the Bose component in the Bose-Fermi
mixture could exhibit
Anderson localization with large exponential tail due to a near
cancellation of the  intra-species
repulsion and inter-species attraction while no localization is
possible in the absence of the OL. No evidence of Anderson localization
with long exponential tail
was found in the Fermi component.  The cigar-shaped Bose-Fermi mixture
is described by an effective one-dimensional GP equation for bosons
coupled to a mean-field hydrodynamic equation for fermions at unitarity
\cite{PRA-81-053630}.
In this study, we use both numerical and variational solution of the
mean-field equation for the mixture.
We obtain a phase plot of number of bosonic and fermionic atoms showing
the domain of symbiotic localization without external trap and
localization in the presence of a quasi-periodic bichromatic OL.
The numerical and variational densities of the localized
states are in good agreement with each other.  Both symbiotic localization
and Anderson localization are found to be dynamically stable when given
a small initial velocity to one of the components. The Bose and Fermi
components are found to move together under such a perturbation.
With present know how, it should be possible to study the
Anderson localization of the Bose component in the Bose-Fermi mixture
on a bichromatic OL experimentally under appropriate condition.

\acknowledgments

FAPESP (Brazil) and CNPq (Brazil) provided partial support.
The Science and Technology Program of the
Education Department of Hubei, China, under Grants No. D200722003 and
Z200722001, provided support during the initial stage of the project in China.

\end{document}